\documentclass{nature}

\usepackage{xspace}
\usepackage{epsfig}
\usepackage{graphicx}
\usepackage{fancyhdr}
\usepackage{pslatex}   
\usepackage{xcolor}
\usepackage{color}
\usepackage{amsmath, amsthm, amssymb} 

\usepackage{lineno}

\newcommand {\lam} {\mbox{$\Lambda$}\xspace}
\newcommand {\lamBar} {\mbox{$\overline{\Lambda}$}\xspace}
\newcommand {\Jsys} {\mbox{$\hat{J}_{\rm sys}$}\xspace}

\newcommand {\RPres} {\mbox{$R^{(1)}_{\rm EP}$}\xspace}
\newcommand {\phiJ} {\mbox{$\phi_{\small{\Jsys}}$}\xspace}
\newcommand{\rootsnn} {\mbox{$\sqrt{s_{\rm NN}}$}\xspace}

\definecolor{dgreen}{cmyk}{1.,0.,1.,0.2}        
\definecolor{orange}{cmyk}{0.,0.353,1.,0.}    

\usepackage{datetime}

\begin{document}

\title{Global $\Lambda$ hyperon polarization in nuclear collisions: evidence for the most vortical fluid}

\maketitle



{\bf 
The extreme temperatures and energy densities generated by
ultra-relativistic collisions between heavy nuclei produce a state
of matter with surprising fluid properties\cite{Adams:2005dq}.
Non-central collisions have angular momentum on the order of $1000\hbar$, and
 the resulting fluid may have a strong vortical
structure\cite{Liang:2004ph,Becattini:2007sr,Pang:2016igs} 
 that must be understood to properly
describe the fluid.  It is also of particular interest
 because the restoration of fundamental symmetries of
quantum chromodynamics is expected to produce novel physical effects
in the presence of strong vorticity\cite{Kharzeev:2015znc}.
However, no experimental indications of fluid vorticity in heavy ion collisions
     have so far been found.
Here we present the first measurement of an alignment between the
angular momentum of a non-central collision and the spin of emitted
particles, revealing that the fluid produced in heavy ion collisions
is by far the most vortical system ever observed.
We find that $\Lambda$ and $\overline{\Lambda}$ hyperons show a
positive polarization of the order of a few percent, consistent with
some hydrodynamic predictions\cite{Becattini:2013vja}.  
A previous measurement\cite{Abelev:2007zk} that reported a null result at higher collision energies
     is seen to be consistent with the trend of our new observations, though with larger statistical uncertainties.
These data provide the first experimental access to the
  vortical structure of the ``perfect fluid''\cite{Heinz:2013th} created in a heavy ion collision.  
They should prove valuable in the development of
  hydrodynamic models
  that quantitatively connect observations
  to the theory of the Strong Force.
Our results extend the recent discovery\cite{FluidSpintronics2015} of
  hydrodynamic spin alignment to the subatomic realm.
}
\vspace*{2mm}


The primary objective of the Relativistic Heavy Ion Collider (RHIC) at
Brookhaven National Laboratory is to produce a large (relative to the
size of a proton) system of matter at temperatures of
$T\approx200~{\rm MeV}/k_B\approx2.3\times10^{12}~{\rm K}$ by
colliding gold nuclei traveling at 96.3 -- 99.995\%
  of the speed of light.  Such temperatures, more than
$100,000$ times that at the Sun's core, characterized the universe
only a few microseconds after the Big Bang\cite{Kolb:1990vq}.
Under these extreme conditions, the protons and neutrons that comprise
our everyday world, melt into a state of deconfined quarks and gluons
called the quark-gluon plasma\cite{Shuryak:1980tp,Adams:2005dq}.  Before RHIC was turned on
in 1999, the expectation was that this plasma would be weakly
coupled and highly viscous.
However, the discovery 
of strong collective behaviour led to the surprising conclusion that
the system generated in these collisions was in fact a liquid with the
lowest viscosity ever observed, the ``nearly perfect
fluid''\cite{Heinz:2013th}.

Since then, large teams have undertaken a program of experimental
  investigation, and increasingly sophisticated hydrodynamic theory has
  proven remarkably successful in reproducing observed properties of the
  fluid\cite{Csernai:2014cwa}.
A complete understanding of this fluid may provide deep insights into
  the strongest and most poorly understood of the fundamental forces in
  nature. 
Quantum chromodynamics (QCD) is the theory of the strong interactions between quarks
     and gluons, but experimental input from RHIC is 
     essential to understand quark confinement and the origin of hadron mass. 

\begin{figure}[t!]
{\centerline{\includegraphics[width=0.7\textwidth]{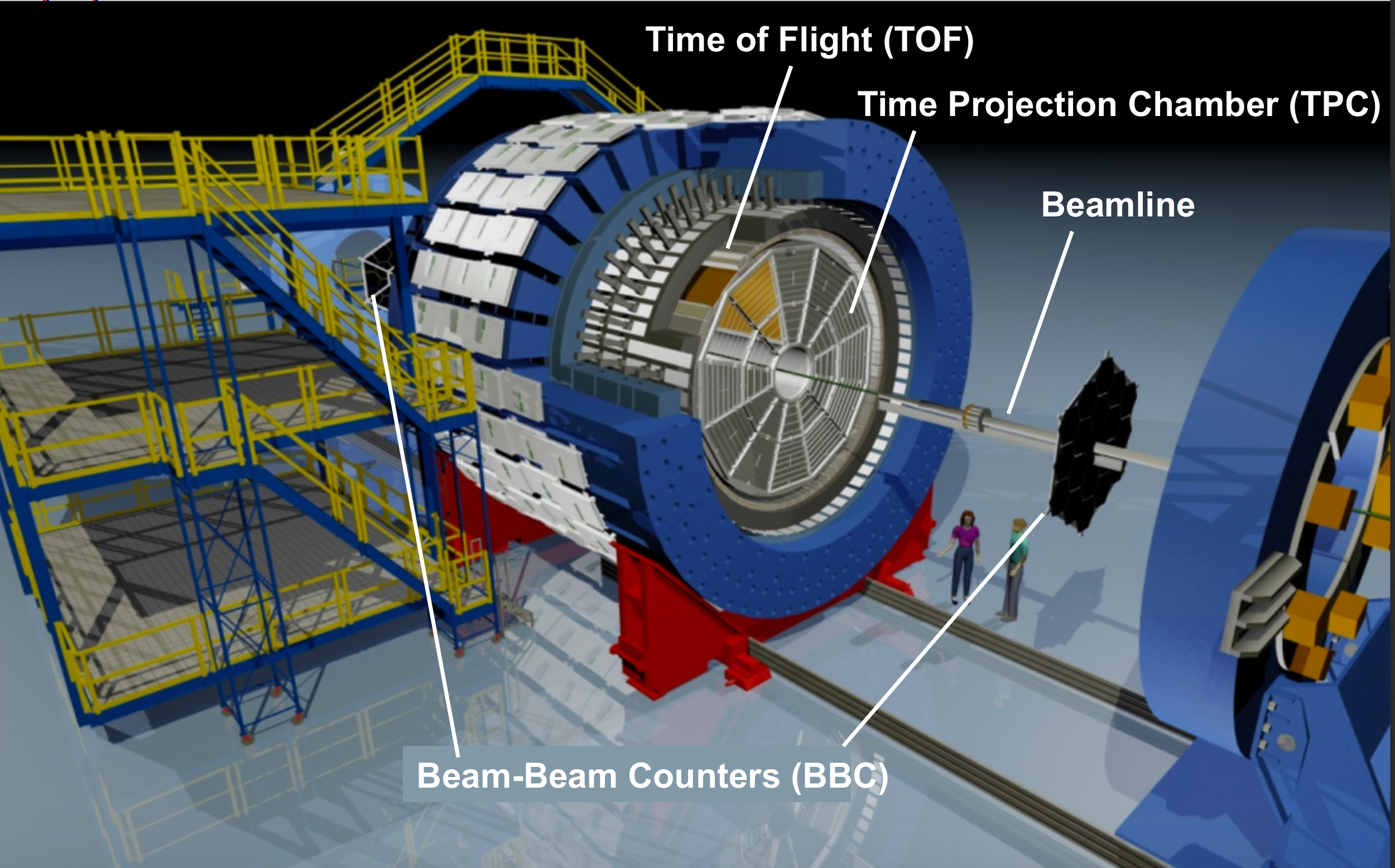}}}
\caption{
\label{fig:SchmahSTAR}
The STAR detector system.  Gold nuclei traveling at nearly the speed
of light travel along the beamline and collide in the center of the
detector system.  Charged particles emitted at midrapidity
(i.e. having a relatively small component of velocity along the beam direction)
are measured in
the Time Projection Chamber (see also figure~\ref{fig:EndOnTPC}) and
the Time-of-Flight detectors.  Forward- and backward-going fragments are detected in
the Beam-Beam Counters.}
\end{figure}

\begin{figure}[t!]
{\centerline{\includegraphics[width=0.6\textwidth]{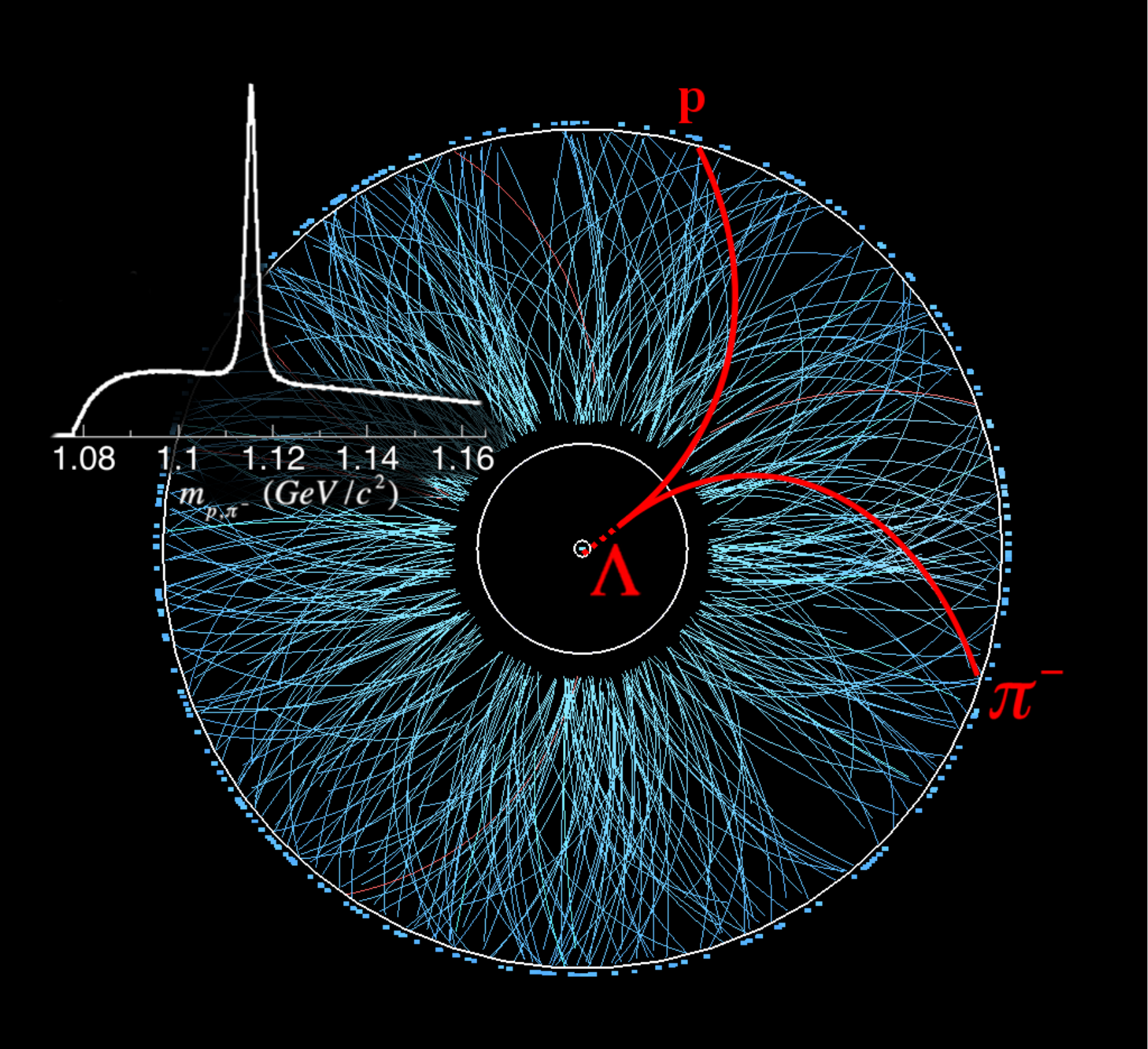}}}
\caption{
\label{fig:EndOnTPC}
Charged particles from a single Au+Au collision ionize the gas in the TPC, forming tracks that curve
in the magnetic field of the detector.
The tracks are reconstructed in three dimensions, making them relatively
  easy to distinguish, but are projected onto a single plane in this figure.
As the tracks
exit at the outer radius, they leave a signal in the Time-of-Flight
(TOF) detector.  The species of charged particles is determined by the
amount of ionization in the TPC and the flight time as measured by
TOF.  Charged daughters from the weak decay $\Lambda\rightarrow{\rm p}+\pi^-$ are extrapolated backwards, and the parent is identified
through topological selection.  A clear peak at the $\Lambda$ mass, obtained by summing over many events, is observed in
the invariant-mass distribution, shown in the inset.}
\end{figure}

A collaboration of physicists from 13 countries operates
the STAR detector system\cite{Ackermann:2002ad} which has recorded
billions of collisions at RHIC.  A rendering of the STAR experiment is
shown in figure~\ref{fig:SchmahSTAR}.  Opposing beams of gold nuclei
collide in the center of the Time Projection Chamber (TPC), generating
a spray of charged particles.
The TPC signal from a single event is shown in
figure~\ref{fig:EndOnTPC}.  Forward- and backward-traveling particles and fragments
that experience only a small deflection are
measured in the Beam-Beam Counters.

\begin{figure}[t!]
{\centerline{\includegraphics[width=0.7\textwidth]{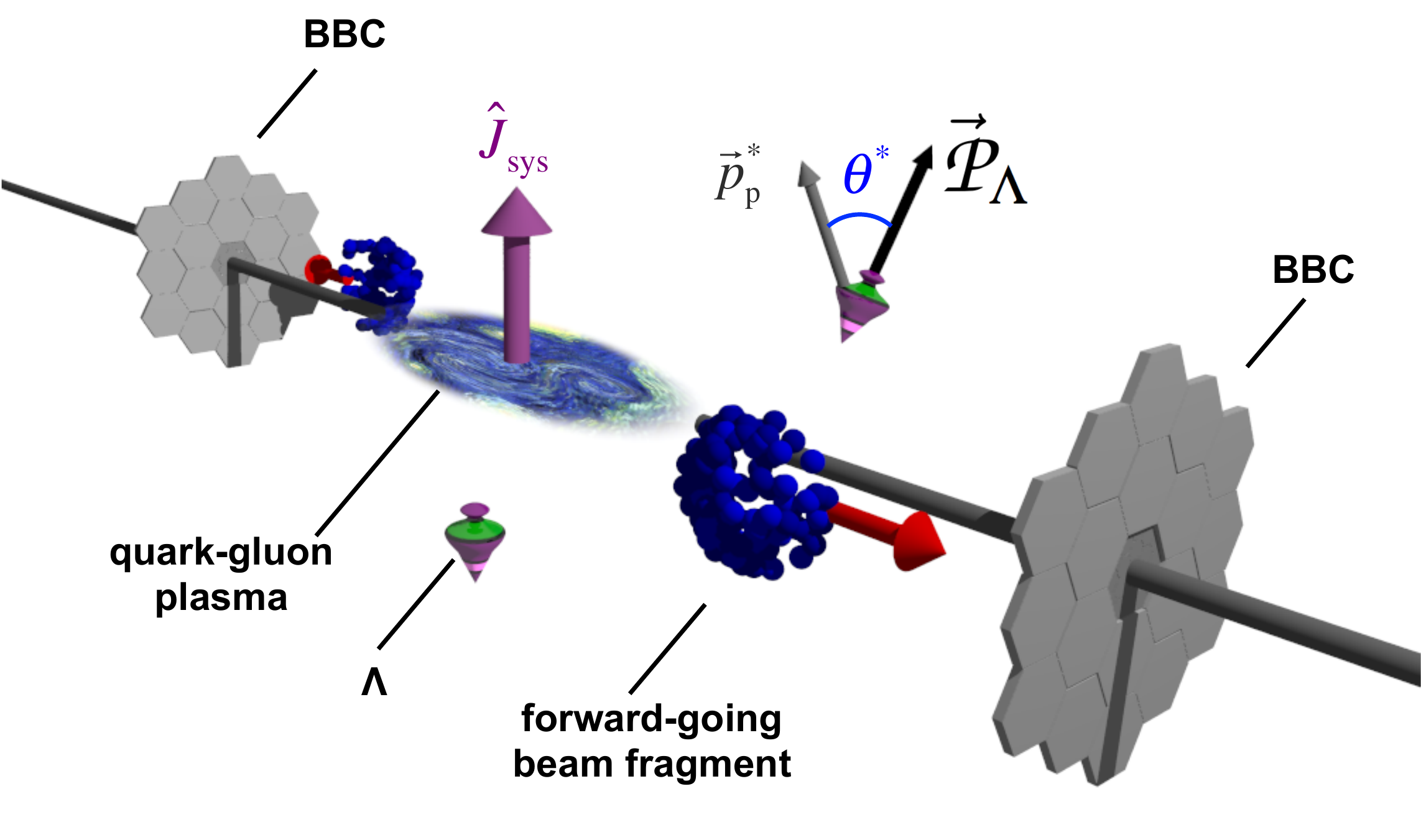}}}
\caption{
\label{fig:Cartoon}
A sketch of the immediate aftermath of a Au+Au collision.  The
vorticity of fluid created at midrapidity is suggested.  
The average vorticity points along the direction of
the angular momentum of the collision, \Jsys.  This direction is
estimated experimentally by measuring the sidewards deflection of the
forward- and backward-going fragments and particles in the BBC detectors.
$\Lambda$ hyperons are depicted as spinning tops; see
text for details.  Obviously, elements in this depiction are not drawn
to scale: the fluid and the beam fragments have sizes
of a few femtometers, whereas the radius of each
BBC is about one meter. 
}
\end{figure}

Most collisions at RHIC are not head-on, and so involve significant
angular momentum -of order $1000\hbar$ for a 
typical collision.
A slight sideward deflection of the forward- and backward-traveling
fragments\cite{Voloshin:2016ppr} from a given collision allows
experimental determination 
of the direction of the overall angular
momentum, \Jsys, as shown schematically in figure~\ref{fig:Cartoon}.

Recently, Takahashi {\it et al.}\cite{FluidSpintronics2015} reported the
first observation of a coupling between the vorticity of a fluid and
the internal quantum spin of the electron, opening the door to a new
field of fluid spintronics.  In their study, vorticity $\vec{\omega}$-- a measure of
the ``swirl'' of the velocity flow field around any point
(non-relativistically,
$\vec{\omega}=\tfrac{1}{2}\,\vec{\nabla}\times\vec{v}$) -- is
generated through shear viscous effects as liquid mercury flows next to a
rigid wall.

In a heavy ion collision, shear forces generated by the
interpenetrating nuclei may present an analogous situation,
introducing vorticity to the fluid.  Indeed, hydrodynamic
calculations predict\cite{Becattini:2015ska} tremendous vorticity in
the fluid at RHIC.  So far, no experimental evidence of vorticity at
RHIC has been reported, and its role in the fluid evolution has not
been explored extensively at the theoretical level.

The vorticity is
currently of intense interest, since it is a key ingredient in
theories that predict observable effects associated with chiral
symmetry restoration and the production of false QCD vacuum
states\cite{Kharzeev:2015znc}.

Spin-orbit coupling can generate a spin alignment,
  or polarization, along the 
  direction of the vorticity which is on average parallel to $\hat{J}_{\rm sys}$\cite{Liang:2004ph,Becattini:2007sr}.
Thus, polarization measurements of hadrons emitted from the fluid can be used to determine 
   $\omega\equiv|\vec{\omega}|$.

It is difficult to measure the spin direction of most hadrons emitted
in a heavy ion collision.  However \lam and \lamBar hyperons are
``self-analyzing.'' That is, in the weak decay $\Lambda\rightarrow
{\rm p}+\pi^-$, the proton tends to be emitted along the spin
direction of the parent \lam\cite{Pondrom:1985}.  
If $\theta^*$ is the angle between the daughter proton
(antiproton) momentum $\vec{p}^*_{\rm p}$ and $\Lambda$ ($\overline{\Lambda}$) polarization vector
$\vec{\mathcal{P}}_{\rm H}$ in the hyperon rest frame, then
\begin{equation}
\label{eq:SelfAnalyzing}
\frac{dN}{d\cos\theta^*} 
= \tfrac{1}{2}\left(1+\alpha_{\rm H}|\vec{\mathcal{P}}_{\rm H}|\cos\theta^*\right) .
\end{equation}
The subscript ${\rm H}$ denotes $\Lambda$ or $\overline{\Lambda}$, and the decay parameter
$\alpha_{\Lambda}=-\alpha_{\overline{\Lambda}}=0.642\pm 0.013$\cite{Agashe:2014kda}.
The angle $\theta^*$ is indicated in figure~\ref{fig:Cartoon}, in
which $\Lambda$ hyperons are depicted as tops spinning about their polarization
direction.

The polarization may depend on the momentum of the emitted hyperons.
However, when averaged over all phasespace, symmetry demands that
$\vec{\mathcal{P}}_{\rm H}$ is parallel to \Jsys.  
Because our limited sample sizes prohibit exploration of these dependences, our analysis
  assumes that $\vec{\mathcal{P}}_{\rm H}$ is independent of momentum, and we extract only
  an average projection of the polarization on \Jsys.
This average may be written\cite{Abelev:2007zk} as
\begin{equation}
\overline{\mathcal{P}}_{\rm H} \equiv \langle\vec{\mathcal{P}}_{\rm H}\cdot\hat{J}_{\rm sys}\rangle =
   \frac{8}{\pi\alpha_{\rm H}}\frac{\left\langle\cos\left(\phi_p^*-\phiJ\right)\right\rangle}{\RPres}, 
\label{eq:MeasuredQuantity}
\end{equation}
where \phiJ is the azimuthal angle of the angular momentum of the
collision, $\phi^*_p$ is the azimuthal angle of the daughter proton
(antiproton) momentum in the $\Lambda$ frame, and \RPres is a factor
that accounts for the finite resolution with which we determine \phiJ\cite{Abelev:2007zk}.
The overline on $\overline{\mathcal{P}}_{\rm H}$ and brackets $\langle\cdots\rangle$
denote an average over 
events and the momenta of $\Lambda$ hyperons detected in the TPC.
Equation~\ref{eq:MeasuredQuantity} is strictly valid only in a perfect detector;
  angle-dependent detection efficiency leads to a correction factor\cite{Abelev:2007zk}
  shifting the results in the present analysis by about 3\%.

A relativistic heavy ion collision can produce several hundred charged particles in our detectors.
For a given energy, a head-on collision produces the maximum number of emitted particles,
       while a glancing one produces only a few.
     To concentrate on collisions with sufficient overlap to produce a fluid with large
       angular momentum, we select events producing an intermediate number of tracks in the TPC.
     Twenty percent of all observed collisions produce more tracks than the collisions studied here,
       while 50\% produce fewer; in the parlance of the field, this is known as a 20-50\% centrality selection.

Equation~\ref{eq:MeasuredQuantity} quantifies an average alignment between hyperon spin and a global feature of
  the collision and is hence a ``global polarization''\cite{Liang:2004ph}.
This is distinct from the well-known phenomenon~ of $\Lambda$ polarization
  at very forward angles in proton-proton collisions\cite{Bunce:1976yb}.
The polarization direction from this latter effect depends on $\Lambda$ momentum and not the global angular momentum;
  it has zero magnitude at midrapidity.

\begin{figure}[t!]
{\centerline{\includegraphics[width=0.6\textwidth]{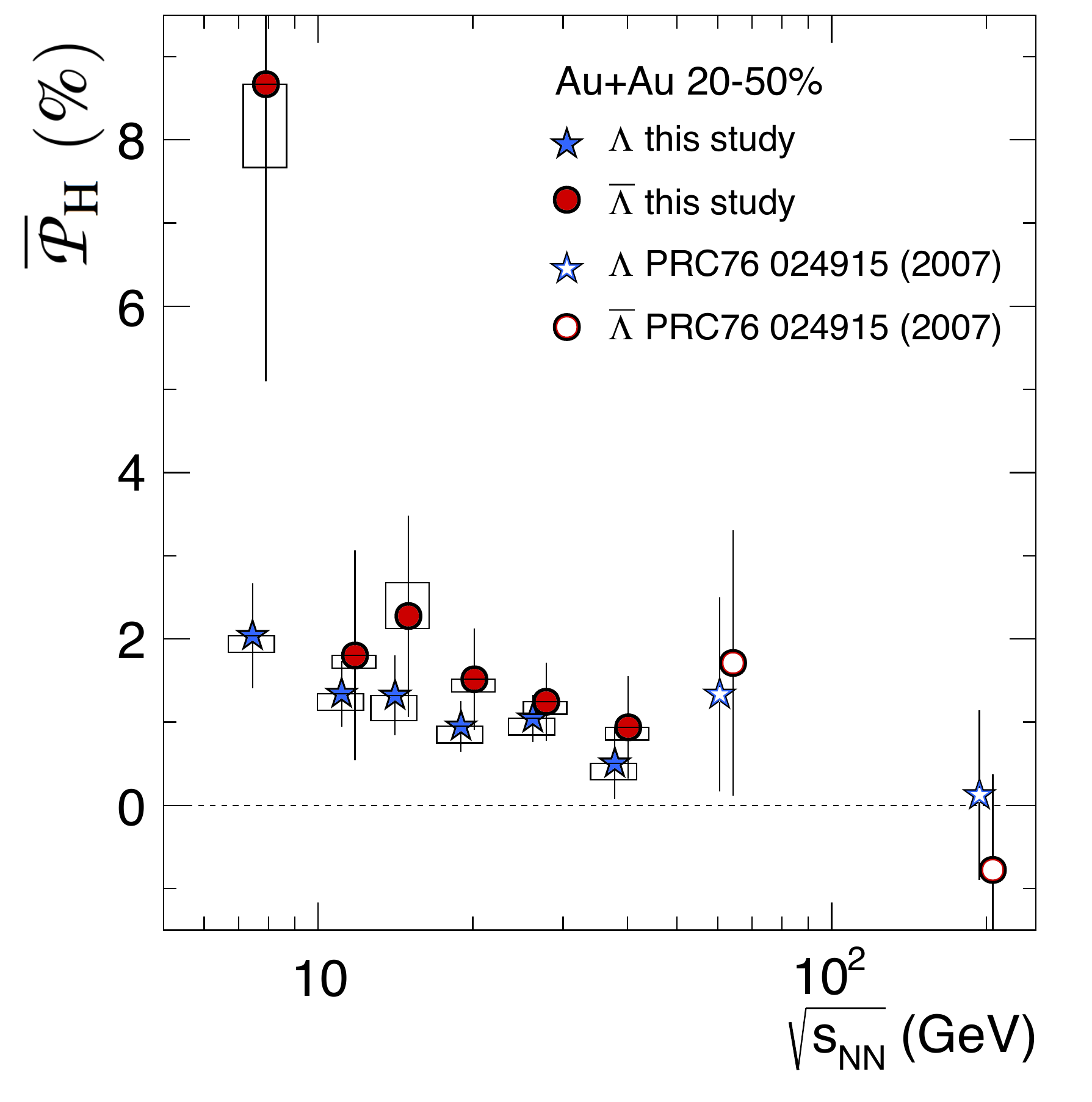}}}
\caption{
\label{fig:Pboth}
The average polarization $\overline{P}_{\rm H}$ (where
${\rm H}=\Lambda$ or $\lamBar$) from 20-50\% central Au+Au collisions
is plotted as a function of collision energy.  The results of the
present study ($\rootsnn<40$~GeV) are shown together with those
reported earlier\cite{Abelev:2007zk} for 62.4 and 200~GeV collisions,
for which only statistical errors are plotted.  Boxes indicate
systematic uncertainties.}
\end{figure}

The solid symbols in figure~\ref{fig:Pboth} show our new measurements
  as a function of collision energy, \rootsnn.
At each energy, a positive polarization at the level of 1.1-3.6 times statistical uncertainty
  is observed for both \lam and \lamBar.
Taken in aggregate, the data are statistically consistent with the hypothesis of
  energy-independent polarizations of
  $1.08\pm0.15$ and $1.38\pm0.30$ percent for \lam and \lamBar, respectively.  
Some models predict that the polarization may decrease with collision energy\cite{Pang:2016igs,Betz:2007kg,Jiang:2016woz}.  
While our data is consistent with such a trend, increased statistics would be required to test these predictions definitively.
Also shown as open symbols in figure~\ref{fig:Pboth} are previously published\cite{Abelev:2007zk} measurements at \rootsnn=62.4~GeV and 200~GeV.
The null result reported in that paper may be seen as consistent with our measurements, within reported
  statistical uncertainty.

Systematic uncertainties are shown as boxes in the figure and are generally smaller than
  statistical ones.
They are dominated by fluctuations in the estimated combinatoric background of
  proton-pion pairs whose invariant mass falls within the $\Lambda$ mass peak, but
  which do not come from $\Lambda$ hyperons.
Uncertainties due to $\Lambda$ identification criteria (such as requirements on the spatial proximity of the proton and $\pi$ daughters)
  are negligible.
There are also small systematic uncertainties in the overall scale, which would scale both the value of $\overline{\mathcal{P}}_{\rm H}$ and
  the statistical uncertainty, thus not affecting the statistical significance of the signal.
This includes the uncertainties in the $\Lambda$ decay parameter $\alpha$ (2\%)\cite{Agashe:2014kda},
  the reaction-plane resolution ($\sim2\%$)\cite{Adamczyk:2014ipa},
  and detector efficiency corrections ($\sim 3.5\%$).

The fluid vorticity may be estimated from the data using the hydrodynamic relation\cite{Becattini:2016gvu}
\begin{equation}
  \label{eq:5author}
  \omega=k_B T\left(\overline{\mathcal{P}}_{\Lambda^\prime}+\overline{\mathcal{P}}_{\overline{\Lambda}^{\prime}}\right)/\hbar ,
\end{equation}
  where $T$ is the temperature of the fluid at the moment when particles are emitted from it.
The subscripts ($\Lambda^\prime$ and $\overline{\Lambda}^\prime$) in equation~\ref{eq:5author} indicate
  that these polarizations are for ``primary'' hyperons emitted directly from the fluid.
However,
  most of the \lam and \lamBar hyperons at these collision energies are
  not primary, but are decay products from heavier particles
  (e.g. $\Sigma^{*,+}\rightarrow\Lambda+\pi^+$), which themselves
  would be polarized by the fluid.
The data in figure~\ref{fig:Pboth} contain both primary and these ``feed-down'' contributions.
At these collision energies, the effect of feed-down is estimated\cite{Becattini:2016gvu} to
  produce only $\sim20\%$ differences between the polarization of ``primary'' and ``all'' hyperons.

The \rootsnn-averaged polarizations indicate a vorticity of $\omega\approx\left(9\pm1\right)\times10^{21}~{\rm s^{-1}}$, with a
systematic uncertainty of a factor of 2, mostly due to uncertainties
in the temperature.
This far surpasses the vorticity of all other known fluids, including
solar subsurface flow ($10^{-7}~{\rm s^{-1}}$)\cite{SolarSubsurfaceVorticity};
  large-scale terrestrial atmospheric patterns ($10^{-7}-10^{-5}~{\rm s^{-1}}$)\cite{Perry_midwesternstreamflow};
  supercell tornado cores ($10^{-1}~{\rm s^{-1}}$)\cite{TornadoVorticity}; 
  the Great Red Spot of Jupiter (up to $10^{-4}~{\rm s^{-1}}$)\cite{JupiterRedSpotVorticity};
  and rotating, heated soap bubbles ($100~{\rm s^{-1}}$) used to model climate change\cite{SoapBubbles2013}.
Vorticities of up to $150~{\rm s^{-1}}$ have been measured in
  turbulent flow in bulk superfluid He-II\cite{SuperfluidHe2Vorticity}, and
Gomez~{\it et al}\cite{Gomez906}
  have recently produced superfluid nanodroplets with $\omega\approx10^{7}~{\rm s^{-1}}$.

Relativistic heavy ion collisions are expected to produce intense magnetic
  fields\cite{Skokov:2009qp} parallel to $\hat{J}_{\rm sys}$.
Coupling between the field and the intrinsic magnetic moments of emitted particles
  may induce a larger polarization for $\overline{\Lambda}$ than $\Lambda$ hyperons\cite{Becattini:2016gvu}.
This is not inconsistent with our observations, but probing the field will require more data 
  to reduce statistical uncertainties as well as potential effects related
  to differences in the measured momenta of $\Lambda$ and $\overline{\Lambda}$ hyperons.

The discovery of global $\Lambda$ polarization in non-central heavy ion
  collisions opens new directions in the study of the hottest, least
  viscous -- and now, most vortical -- fluid ever produced in the laboratory.
Quantitative estimates of extreme vorticity
  yield a more complete characterization of the system
  and are crucial input to studies of novel phenomena
  related to chiral symmetry restoration
  that may provide needed insight into the complex interactions between quarks and gluons.

\noindent{\bf Acknowledgements}
We thank the RHIC Operations Group and RCF at BNL, the NERSC Center at LBNL, and the Open Science Grid
consortium for providing resources and support. This work was supported in part by the Office of Nuclear 
Physics within the U.S. DOE Office of Science, the U.S. National Science Foundation, the Ministry of 
Education and Science of the Russian Federation, National Natural Science Foundation of China, Chinese 
Academy of Science, the Ministry of Science and Technology of China and the Chinese Ministry of Education, 
the National Research Foundation of Korea, GA and MSMT of the Czech Republic, Department of Atomic Energy 
and Department of Science and Technology of the Government of India; the National Science Centre of Poland, 
National Research Foundation, the Ministry of Science, Education and Sports of the Republic of Croatia, and
RosAtom of Russia.

\clearpage

\author{
L.~Adamczyk$^{1}$,
J.~K.~Adkins$^{19}$,
G.~Agakishiev$^{17}$,
M.~M.~Aggarwal$^{31}$,
Z.~Ahammed$^{50}$,
N.~N.~Ajitanand$^{40}$,
I.~Alekseev$^{15,26}$,
D.~M.~Anderson$^{42}$,
R.~Aoyama$^{46}$,
A.~Aparin$^{17}$
% ...
}
\\
\\ (STAR Collaboration)
\\
\normalsize{$^{1}$AGH University of Science and Technology, FPACS, Cracow 30-059, Poland}\\
\normalsize{$^{2}$Argonne National Laboratory, Argonne, Illinois 60439}\\
\normalsize{$^{3}$Brookhaven National Laboratory, Upton, New York 11973}\\
\normalsize{$^{4}$University of California, Berkeley, California 94720}\\
\normalsize{$^{5}$University of California, Davis, California 95616}\\
\normalsize{$^{6}$University of California, Los Angeles, California 90095}\\
\normalsize{$^{7}$Central China Normal University, Wuhan, Hubei 430079}\\
\normalsize{$^{8}$University of Illinois at Chicago, Chicago, Illinois 60607}\\
\normalsize{$^{9}$Creighton University, Omaha, Nebraska 68178}\\
\normalsize{$^{10}$Czech Technical University in Prague, FNSPE, Prague, 115 19, Czech Republic}\\
\normalsize{$^{11}$Nuclear Physics Institute AS CR, 250 68 Prague, Czech Republic}\\
\normalsize{$^{12}$Frankfurt Institute for Advanced Studies FIAS, Frankfurt 60438, Germany}\\
\normalsize{$^{13}$Institute of Physics, Bhubaneswar 751005, India}\\
\normalsize{$^{14}$Indiana University, Bloomington, Indiana 47408}\\
\normalsize{$^{15}$Alikhanov Institute for Theoretical and Experimental Physics, Moscow 117218, Russia}\\
\normalsize{$^{16}$University of Jammu, Jammu 180001, India}\\
\normalsize{$^{17}$Joint Institute for Nuclear Research, Dubna, 141 980, Russia}\\
\normalsize{$^{18}$Kent State University, Kent, Ohio 44242}\\
\normalsize{$^{19}$University of Kentucky, Lexington, Kentucky, 40506-0055}\\
\normalsize{$^{20}$Lamar University, Physics Department, Beaumont, Texas 77710}\\
\normalsize{$^{21}$Institute of Modern Physics, Chinese Academy of Sciences, Lanzhou, Gansu 730000}\\
\normalsize{$^{22}$Lawrence Berkeley National Laboratory, Berkeley, California 94720}\\
\normalsize{$^{23}$Lehigh University, Bethlehem, PA, 18015}\\
\normalsize{$^{24}$Max-Planck-Institut fur Physik, Munich 80805, Germany}\\
\normalsize{$^{25}$Michigan State University, East Lansing, Michigan 48824}\\
\normalsize{$^{26}$National Research Nuclear University MEPhI, Moscow 115409, Russia}\\
\normalsize{$^{27}$National Institute of Science Education and Research, Bhubaneswar 751005, India}\\
\normalsize{$^{28}$National Cheng Kung University, Tainan 70101 }\\
\normalsize{$^{29}$Ohio State University, Columbus, Ohio 43210}\\
\normalsize{$^{30}$Institute of Nuclear Physics PAN, Cracow 31-342, Poland}\\
\normalsize{$^{31}$Panjab University, Chandigarh 160014, India}\\
\normalsize{$^{32}$Pennsylvania State University, University Park, Pennsylvania 16802}\\
\normalsize{$^{33}$Institute of High Energy Physics, Protvino 142281, Russia}\\
\normalsize{$^{34}$Purdue University, West Lafayette, Indiana 47907}\\
\normalsize{$^{35}$Pusan National University, Pusan 46241, Korea}\\
\normalsize{$^{36}$Rice University, Houston, Texas 77251}\\
\normalsize{$^{37}$University of Science and Technology of China, Hefei, Anhui 230026}\\
\normalsize{$^{38}$Shandong University, Jinan, Shandong 250100}\\
\normalsize{$^{39}$Shanghai Institute of Applied Physics, Chinese Academy of Sciences, Shanghai 201800}\\
\normalsize{$^{40}$State University Of New York, Stony Brook, NY 11794}\\
\normalsize{$^{41}$Temple University, Philadelphia, Pennsylvania 19122}\\
\normalsize{$^{42}$Texas A\&M University, College Station, Texas 77843}\\
\normalsize{$^{43}$University of Texas, Austin, Texas 78712}\\
\normalsize{$^{44}$University of Houston, Houston, Texas 77204}\\
\normalsize{$^{45}$Tsinghua University, Beijing 100084}\\
\normalsize{$^{46}$University of Tsukuba, Tsukuba, Ibaraki, Japan,}\\
\normalsize{$^{47}$Southern Connecticut State University, New Haven, CT, 06515}\\
\normalsize{$^{48}$United States Naval Academy, Annapolis, Maryland, 21402}\\
\normalsize{$^{49}$Valparaiso University, Valparaiso, Indiana 46383}\\
\normalsize{$^{50}$Variable Energy Cyclotron Centre, Kolkata 700064, India}\\
\normalsize{$^{51}$Warsaw University of Technology, Warsaw 00-661, Poland}\\
\normalsize{$^{52}$Wayne State University, Detroit, Michigan 48201}\\
\normalsize{$^{53}$World Laboratory for Cosmology and Particle Physics (WLCAPP), Cairo 11571, Egypt}\\
\normalsize{$^{54}$Yale University, New Haven, Connecticut 06520}

\end{document}